\documentclass[reprint,noshowpacs,noshowkeys,prd,balancelastpage,nofootinbib]{revtex4}
\usepackage{amsfonts}
\usepackage{mdframed}
\usepackage{amssymb}
\usepackage{footnote}
\usepackage{amsmath}
\usepackage{graphicx}
\usepackage{float}
\usepackage[font={footnotesize,it}]{caption}
\usepackage[utf8]{inputenc}
\usepackage{natbib}
\usepackage{tikz}
\usepackage{tikz-3dplot}
\usepackage[colorlinks=true,
            linkcolor=purple,
            urlcolor=purple,
            citecolor=blue]{hyperref}

\begin{document}

\title{A Closed Universe: de Sitter Cosmic Gate}
\author{S. Danial Forghani}
\email{danial.forghani@final.edu.tr}
\affiliation{Faculty of Engineering, Final International University, Kyrenia, North
Cyprus via Mersin 10, Turkey}
\author{S. Habib Mazharimousavi}
\email{habib.mazhari@emu.edu.tr}
\affiliation{Department of Physics, Faculty of Arts and Sciences, Eastern Mediterranean
University, Famagusta, North Cyprus via Mersin 10, Turkey}
\date{\today }

\begin{abstract}
A new cosmological object in analogy with the concept of
a wormhole in general relativity is introduced. As wormholes connect two distant points
through a tunnel in a spacetime, this new object connects two
spacetime through a large mouth which is referred to as a "Cosmic Gate". In this
context, two identical copies of the regular part of the de Sitter spacetime are cut through a timelike hyperplane, and are glued at their identical boundaries (the only boundary) to form a complete spacetime. The stability of the cosmic gate against a linear radial perturbation is studied as well. Finally, the radial geodesics of the spacetime for a timelike and a null particle are presented.
\end{abstract}

\maketitle

\section{Introduction}

In his 1985 hard sci-fi novel "Contact", Carl Sagan portraits a spacetime
journey via a wormhole in the most scientific way possible for its time. In
his endeavor to build up such an accurate presentation of wormholes for a
novel, he benefited from Kip Thorne's consultant who at the time was a
physics professor at Caltech. Later in 1987, the communication between the
two scientists initiated the idea of two scientific masterworks, published
in 1988 by Morris and Thorne \cite{1,2}. The significance of these articles
lies in the fact that, although wormholes and their application for space
and time travel were known by 1988, they were not considered to be
practically traversable. Therefore, the ever-fainting trend of wormhole
physics since the 1950s was considerably revived, which in turn led to many
other great papers in the followed by years \cite{3,4,5,6,7,8,9,10}.
However, it was already shown by Morris and Thorne {\cite{2}} that in order
to keep the throat of a wormhole open, matter with negative energy density
had to be distributed over the spacetime. In 1989, Matt Visser, inspired by
thin-shell formalism, constructed a new class of traversable wormholes -
known as thin-shell wormholes - which had this exotic matter\ limited only
to the location of the throat {\cite{3}}. Six years later, Poisson and
Visser proposed a method by which the stability of such wormholes could be
somehow determined under a linear radial perturbation {\cite{11}}.

In this study, we would like to employ the idea of thin-shell wormholes and
apply it to construct a new entity that represents a closed and complete
universe. In thin-shell wormhole studies, it is usual to cut two exterior
solutions in a way that the existed singularities and their associated event
horizons are excised out. Then the two remaining manifolds (that are
geodesically incomplete) are identified at the cut hypersurface to form a
geodesically complete manifold. Of course, the identification of the two
manifolds requires certain conditions to be satisfied which results in
determining the surface energy density and pressure of the matter that
supports the identification location, known as the throat of the wormhole.
In this paper, we instead excise out the exterior solutions and keep the
interior ones to construct a closed and complete universe by gluing them. In
this regard, we must choose a solution that is free of any singularities
and/or horizons at or towards the center at $r=0$. Evidently, the de Sitter
solution {\cite{12,13,14}} perfectly fits the situation. The de Sitter
solution is a maximally symmetric vacuum solution in general relativity that
describes a $3+1$-dimensional expanding universe with a positive
cosmological constant. The main significance of the de Sitter solution is
that our universe is believed to be asymptotically de Sitter, based on
cosmological observations {\cite{15}}.

The paper is arranged as follows. In section {II}, we briefly discuss
wormholes in Morris and Thorne's sense to review some of their important
features related to our discussion. Section {III} goes into the details of
our construction. We show that the resultant thin-shell is supported by
ordinary matter and is stable under a linear radial perturbation. In the
continuation, we show in section {IV} that our constructed close universe is
also geodesically complete. We conclude our paper in section {V}.
All over the paper we have used the conventional units $c=G=1$.

\section{Wormhole vs. Gate}

The so-called Morris-Thorne wormhole spacetime is described by 
\begin{equation}
ds^{2}=-e^{2\Phi \left( r\right) }dt^{2}+\frac{dr^{2}}{1-\frac{b\left(
r\right) }{r}}+r^{2}d\Omega ^{2},  \label{1}
\end{equation}%
in which the red-shift function $e^{2\Phi \left( r\right) }$ admits no root
in the domain of $r$ and $\lim_{r\rightarrow \infty }\Phi \rightarrow 0$.
Furthermore, the shape function $b\left( r\right) $ satisfies the so-called 
\textit{flare-out conditions }\cite{16}. These conditions are obtained
through applying the embedding process to (\ref{1}), which leads to an
embedding diagram for the wormhole. The embedding diagram is to visualize a $%
2$-dimensional version of a $3+1$-dimensional universe by considering an
instant of time ($t=t_{0}$) on the equator ($\theta =\pi /2$), that reduces
the original line element in (\ref{1}) to 
\begin{equation}
ds^{2}=\frac{dr^{2}}{1-\frac{b\left( r\right) }{r}}+r^{2}d\phi ^{2}.
\label{2}
\end{equation}%
Then, we embed this $2$-dimensional surface onto a $3$-dimensional flat
cylindrical space by comparing it to the line element 
\begin{equation}
ds^{2}=dr^{2}+dz^{2}+r^{2}d\phi ^{2}.  \label{3}
\end{equation}%
With the cylindrical symmetry configuration, we assume $z=z\left( r\right) $%
, which takes us to 
\begin{equation}
\frac{1}{1-\frac{b\left( r\right) }{r}}=1+z^{\prime }\left( r\right) ^{2},
\label{4}
\end{equation}%
where $z^{\prime }\left( r\right) =\frac{dz\left( r\right) }{dr}$.
Consequently, we obtain 
\begin{equation}
z\left( r\right) =\pm \int_{R_{0}}^{r}\frac{1}{\sqrt{\frac{r}{b\left(
r\right) }-1}}dr,  \label{5}
\end{equation}%
in which at $r=R_{0}$ we assume $z\left( r\right) =z\left( R_{0}\right) =0.$
Accordingly, $z\left( R_{0}\right) $ is called the throat of the wormhole
connecting the region $z\geq 0$ to the region $z\leq 0$ provided that $%
r\left( z\right) $ is the absolute minimum at $r\left( 0\right) =R_{0}.$
Technically speaking, this condition means that $b\left( R_{0}\right) =R_{0}$
and $b-rb^{\prime }\geq 0$ for $r\geq R_{0}$. In order to keep the signature
of the spacetime unaltered for $r\geq R_{0}$, one should also add the
condition $b\left( r\right) <r$ to the previous conditions. Therefore, while
in the embedding diagram we have $z\in \mathbb{R}$, the condition $r\geq
R_{0}$ always holds. One can mathematically prove that the requirement of
these conditions, i.e. the flare-out conditions, implies that the
energy-momentum tensor supporting the wormhole spacetime does not satisfy
the null energy condition, hence the corresponding matter is exotic \cite{2}.

Next, let us assume, there exists a spacetime described by the following
line element 
\begin{equation}
ds^{2}=-e^{2\Phi \left( r\right) }dt^{2}+\frac{dr^{2}}{\frac{b\left(
r\right) }{r}-1}+r^{2}d\Omega ^{2},  \label{6}
\end{equation}%
such that $b\left( r\right) \geq r$ for $r\leq R_{0}$ and the equality
occurs at $r=R_{0}.$ Here, $r=R_{0}$, is not a throat but in the sequel we
shall refer to it as a \textquotedblleft \textit{Gate}\textquotedblright .
Looking at its embedding diagram, we follow the steps as in the case for the
Morris-Thorne wormhole. Hence, we obtain 
\begin{equation}
\frac{1}{\frac{b\left( r\right) }{r}-1}=1+z^{\prime }\left( r\right) ^{2},
\label{7}
\end{equation}%
and as a result 
\begin{equation}
z^{\prime }\left( r\right) =\pm \sqrt{\frac{2-\frac{b\left( r\right) }{r}}{%
\frac{b\left( r\right) }{r}-1}.}  \label{8}
\end{equation}%
Having $z^{\prime }\left( r\right) $ real, one has to impose the condition 
\begin{equation}
1\leq \frac{b}{r}\leq 2.  \label{9}
\end{equation}%
We observe that, at $r=R_{0}$, $\frac{dr}{dz}=\frac{1}{z^{\prime }}$ is
still zero, however, its second derivative is negative, i.e. 
\begin{equation}
\frac{d^{2}r}{dz^{2}}<0.  \label{10}
\end{equation}%
This implies that $r\left( z\right) $ at $z=0$ becomes an absolute maximum.
The function $z\left( r\right) $ is obtained to be 
\begin{equation}
z\left( r\right) =\pm \int_{R_{0}}^{r}\sqrt{\frac{2-\frac{b\left( r\right) }{%
r}}{\frac{b\left( r\right) }{r}-1},}  \label{11}
\end{equation}%
where we again considered $z\left( R_{0}\right) =0$ and $z\in \lbrack
-R_{0},R_{0}].$ On this account, unlike the wormhole spacetime, here the
gate connects two bounded/closed spacetimes at their maximum surface. The
resultant spacetime is complete and consists of two parts connected by what
we are calling a \textquotedblleft \textit{Cosmic Gate}\textquotedblright\
from here onwards, in the hope that the terminology pleases the reader both
scientifically and phonaesthetically.

\section{de Sitter thin-shell cosmic gate}

Let us start by setting the line element of a $3+1$-dimensional spherically
symmetric universe 
\begin{equation}
ds^{2}=-f\left( r\right) dt^{2}+\frac{1}{f\left( r\right) }%
dr^{2}+r^{2}\left( d\theta ^{2}+\sin ^{2}\theta d\phi ^{2}\right) ,
\label{12}
\end{equation}%
which represents the maximally symmetric de Sitter spacetime with a positive
scalar curvature for 
\begin{equation}
f\left( r\right) =1-\frac{r^{2}}{\ell ^{2}},  \label{13}
\end{equation}%
where $\ell ^{2}=\frac{3}{\Lambda }$, with $\Lambda >0$ being the
cosmological constant. The spacetime is nonsingular, although it has a
cosmological horizon at $r_{c}=\ell $. In what follows we construct a closed
and geodesically complete universe by virtue of Visser's cut-and-paste
method originally used for constructing thin-shell wormholes \cite{3}.
However, what eventually results in our consideration is essentially a
thin-shell cosmic gate, as we will show here. Hence, in the bulk metric (\ref%
{12}) we make two identical copies of the spacetime inside a timelike
hypersurface defined by $\Sigma :r=R<r_{c}$, and paste them at this common
hypersurface $\Sigma $ to construct a closed universe. To show the closeness
of this universe, we rely on its embedding diagram, which is shown in Fig. 1
for arbitrary values of $r_{c}$ and $R<r_{c}$. 
\begin{figure}[tbp]
\includegraphics[width=70mm,scale=0.5]{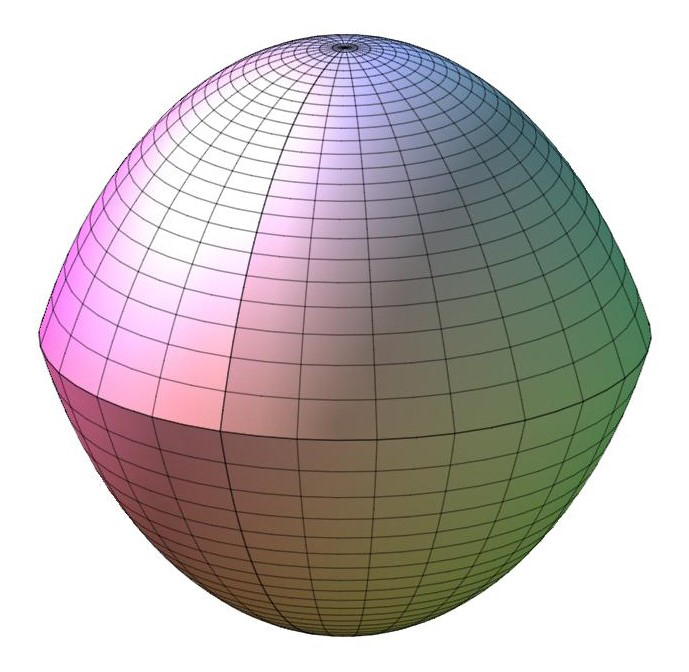}
\caption{Embedding diagram of two identical de-Sitter universes joined at
the cosmic gate. The result is a non-singular closed universe that is
geodesically complete.}
\end{figure}
The function regarding the diagram displayed in Fig. 1 is given by 
\begin{equation}
z=\pm \ell \,\left[ \sqrt{1-\frac{r^{2}}{\ell ^{2}}}-\sqrt{1-\frac{R^{2}}{%
\ell ^{2}}}\right] ,  \label{14}
\end{equation}%
following the method described in section II. Once again, we emphasize that
the flare-out conditions for thin-shell wormholes \cite{17} behave
absolutely inversely in the case of thin-shell cosmic gates to satisfy $%
\frac{d^{2}r}{dz^{2}}=-\frac{\ell ^{2}}{r^{3}}<0$, in our case.

It is known from Darmois-Israel formalism that the boundary hypersurface $%
\Sigma $ should satisfy certain junction conditions \cite{18}. The details
of the procedure are explained in many references \cite{19,20,21,22}, so
here we refrain from excessive repetition. In short, the first fundamental
form (the induced metric tensor) has to be continuous across the thin-shell
whereas the second fundamental form (the induced curvature tensor) is
discontinuous due to the presence of matter at the shell. Satisfying these
conditions leads to mathematical expressions for the energy density $\sigma $
and angular pressure $p$ of the matter present on the shell. The symmetry of
the bulk and the hyperplane $\Sigma $ yields the matter on the shell to be a
perfect fluid with the surface energy-momentum tensor given by $%
S_{a}^{b}=diag\left( -\sigma ,p,p\right) $. We should comment that there is
a fundamental difference between a thin-shell wormhole and a thin-shell
cosmic gate in this context, that lies in the direction of the $4$-normal of
the hypersurface $\Sigma $ - which are actually in opposite directions.
Eventually, the expressions for the energy density and angular pressure read 
\begin{equation}
\sigma =\frac{1}{2\pi R}\,\sqrt{f\left( R\right) +\dot{R}^{2}},  \label{15}
\end{equation}%
and 
\begin{equation}
p=-\frac{1}{4\pi }\,\frac{\ddot{R}+f^{\prime }\left( R\right) /2}{\sqrt{%
f\left( R\right) +\dot{R}^{2}}}-\frac{\sigma }{2},  \label{16}
\end{equation}%
respectively. Here, $\left. \dot{R}\equiv \frac{dr}{d\tau }\right\vert
_{r=R} $ is the speed of the shell measured by an observer on the shell, and 
$\left. f^{\prime }\left( R\right) \equiv \frac{df\left( r\right) }{dr}%
\right\vert _{r=R}$ is the total radial derivative of $f\left( r\right) $ at
the shell. Note that due to the opposite direction of the $4$-normals
explained earlier, the signs of the energy density $\sigma $ and angular
pressure $p$ are opposite the signs of their counterparts in thin-shell
wormhole theory. It is evident from (\ref{15}) that in static equilibrium
configuration where $\ddot{R}=\dot{R}=0$, the weak energy condition \cite{23}
is satisfied for the perfect fluid at the gate i.e., $\sigma \geq 0$ and $%
\sigma +p\geq 0$. Therefore unlike for a wormhole the matter on the
shell/gate is ordinary. Besides, it can be shown by direct substitution that
the relation 
\begin{equation}
\frac{d\sigma }{dR}=-\frac{2}{R}\,\left( \sigma +p\right) ,  \label{17}
\end{equation}%
illustrative of conservation of energy, holds for the matter at the gate.
Performing a second derivative on the latter equation results in 
\begin{equation}
\frac{d^{2}\sigma }{dR^{2}}=\frac{2}{R^{2}}\,\left( \sigma +p\right) \left(
3+2\beta ^{2}\right) +\frac{2\gamma }{R},  \label{18}
\end{equation}%
in which $\beta ^{2}\equiv \frac{\partial p}{\partial \sigma }$, and $\gamma
\equiv -\frac{\partial p}{\partial R}$, provided that a variable equation of
state $p=p\left( \sigma ,R\right) $ \cite{24} governs. Eq. (\ref{15}) could
be written in an alternative form to recollect a classical equation of
motion 
\begin{equation}
\dot{R}^{2}+V_{e}\left( R\right) =0,  \label{19}
\end{equation}%
where 
\begin{equation}
V_{e}\left( R\right) =f\left( R\right) -\left( 2\pi R\sigma \right) ^{2},
\label{20}
\end{equation}%
is the effective potential. In static equilibrium of thin-shell wormholes 
\cite{11}, the first derivative of the effective potential i.e., $%
V_{e}^{\prime }\left( R_{0}\right) $ becomes zero at the supposed
equilibrium radius $R_{0}$, where the sign of the second derivative of the
potential $V_{e}^{\prime \prime }\left( R_{0}\right) $ determines whether
the equilibrium is stable or not. Although in this context what we are
confronting is not a thin-shell wormhole, technically speaking, the same
method could be applied consistently. To this, one should benefit from Eqs. (%
\ref{17}) and (\ref{18}) to arrive at an expression for $V_{e}^{\prime
\prime }\left( R_{0}\right) $. Then, by setting $V_{e}^{\prime \prime
}\left( R_{0}\right) =0$ and solving for $\beta ^{2}$ we obtain 
\begin{equation}
\beta ^{2}=-\frac{1+4\pi \gamma R_{0}^{2}\,f^{3/2}\left( R_{0}\right) }{%
2f\left( R_{0}\right) },  \label{21}
\end{equation}%
where $f\left( R_{0}\right) =1-\frac{R_{0}^{2}}{\ell ^{2}}$. The purpose
here is to graph $\beta ^{2}$ versus the equilibrium radius $R_{0}$ to spot
the stability regions for which $V_{e}^{\prime \prime }\left( R_{0}\right)
>0 $. The parameter $\beta ^{2}$ is physically interpreted as the speed of
sound in matter \cite{11}. In this regard, we are particularly interested in
the regions where $0<\beta ^{2}<1$ since we have set $c=1$. As can also be
observed from Fig. 2, $\beta ^{2}$ is negative-definite for a barotropic
equation of state where $\gamma =0$. On the other hand, for negative values
of $\gamma $, the value of $\beta ^{2}$ may fall into the interval $\left(
0,1\right) $ for stable situations. In Fig. 2, $\beta ^{2}$ is graphed
versus $R_{0}$ for $\ell =1$ and given values of $\gamma $. 
\begin{figure}[tbp]
\includegraphics[width=90mm,scale=1]{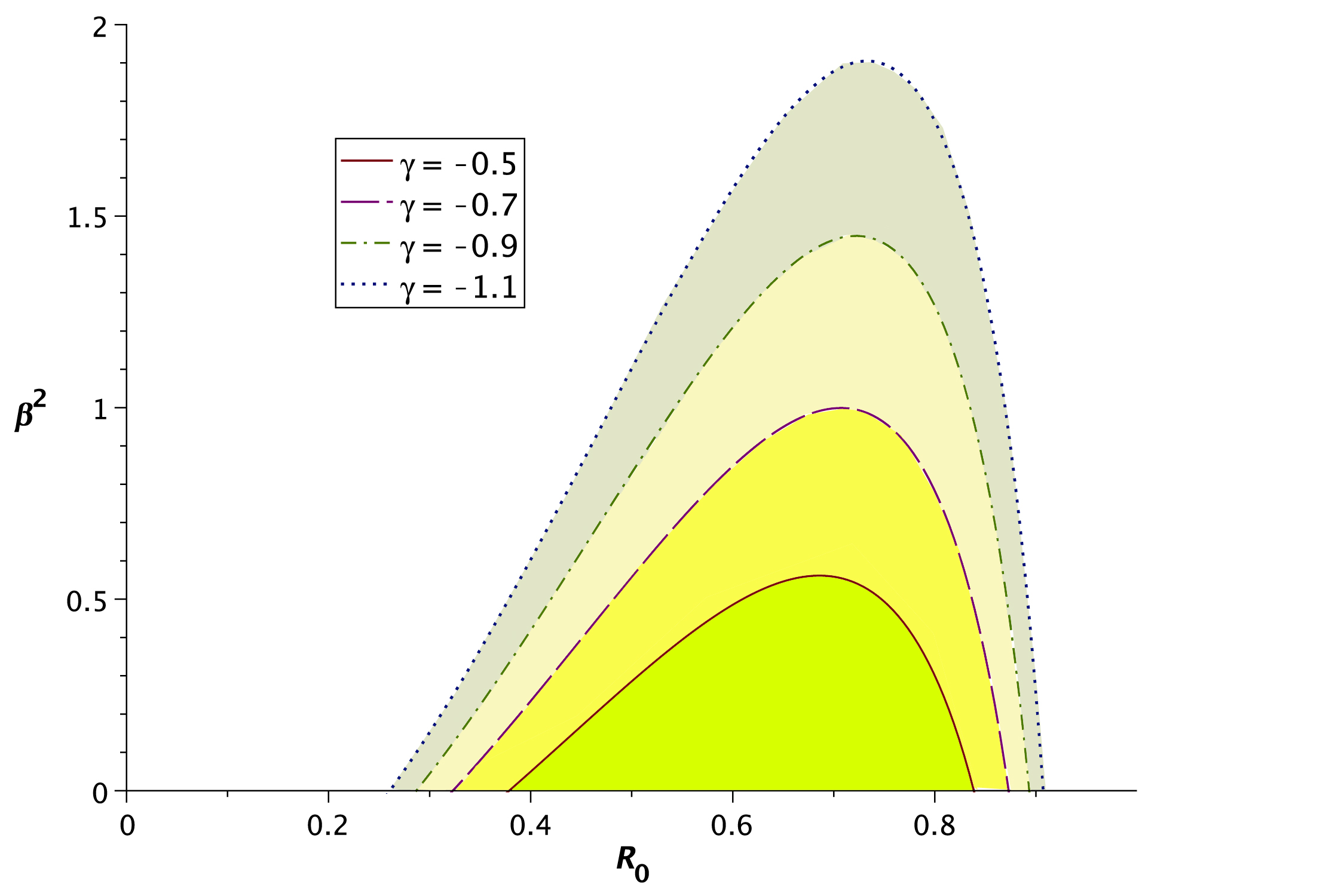}
\caption{The graph for speed of sound $\protect\beta ^{2}$ versus the
equilibrium radius $R_{0}$ for $\protect\alpha =1$ and four given values of $%
\protect\gamma $. The stable regions are shown in colors.}
\end{figure}

\section{The Geodesics}

In this section, we study the geodesics of the constructed closed universe.
In this regard, we are interested in investigating the radial timelike and
null geodesics for which the angular momentum of the test particles is zero.
The Lagrangian of a particle moving on its equatorial plane along its
geodesics in a static spherically symmetric space-time is given by \cite{25} 
\begin{equation}
\mathcal{L}=\frac{1}{2}\left[ -f\left( r\right) \dot{t}^{2}+\frac{1}{f\left(
r\right) }\dot{r}^{2}+r^{2}\dot{\phi}^{2}\right]  \label{22}
\end{equation}%
where an overhead dot represents differentiation with respect to an
arbitrary affine parameter. Furthermore the 4-velocity satisfies $g_{\mu \nu
}\dot{x}^{\mu }\dot{x}^{\nu }=\epsilon $ or explicitly%
\begin{equation}
-f\left( r\right) \dot{t}^{2}+\frac{1}{f\left( r\right) }\dot{r}^{2}+r^{2}%
\dot{\phi}^{2}=\epsilon  \label{23}
\end{equation}%
with $\epsilon =-1$ ($\epsilon =0$) for a timelike (null) geodesics. In the
first observation, since $\mathcal{L}$ is independent of $t$ and $\phi $ the
energy 
\begin{equation}
E\equiv -\frac{\partial \mathcal{L}}{\partial \dot{t}}=f\left( r\right) \dot{%
t}  \label{24}
\end{equation}%
and the angular momentum 
\begin{equation}
L\equiv \frac{\partial \mathcal{L}}{\partial \dot{\phi}}=r^{2}\dot{\phi}
\label{25}
\end{equation}%
are conserved. By substituting the energy and angular momentum into Eq. (\ref%
{23}) one obtains the differential equation governing the radial coordinate
of the geodesics expressed by 
\begin{equation}
\dot{r}^{2}=E^{2}+f\left( r\right) \left( \epsilon -\frac{L^{2}}{r^{2}}%
\right) .  \label{26}
\end{equation}%
In what follows, we consider both timelike ($\epsilon =-1$) and null ($%
\epsilon =0$) cases for a radial motion with no angular momentum ($L=0$).

\subsection{The Timelike Geodesics}

For $\epsilon =-1$, once we set $L=0$ in Eq. (\ref{26}) it reduces to 
\begin{equation}
\frac{dr}{d\tau }=\pm \sqrt{E^{2}-1+\frac{r^{2}}{\ell ^{2}}},  \label{27}
\end{equation}%
where we considered the affine parameter to be the proper time. The latter
equation of motion is exactly solvable with the following solution 
\begin{equation}
r=r_{0}\cosh \left( \pm \frac{\left( \tau -\tau _{0}\right) }{r_{c}}\right) ,
\label{28}
\end{equation}%
where $r_{0}=\ell \sqrt{1-E^{2}}$ ($E^{2}<1$) is the initial radial position
provided that the particle starts from the rest at $\tau =\tau _{0}$.
Naturally, we assume that $r_{0}<R_{0}$. Solving for $r=R_{0}$ shows that
the particle reaches the cosmic gate at ($\tau _{0}=0$) 
\begin{equation}
\tau _{R}=r_{c}\cosh ^{-1}\left( \frac{R_{0}}{r_{0}}\right) .  \label{29}
\end{equation}%
This, however, is measured by the commoving observer, not by a distant
observer. For the distant observer, one needs to solve 
\begin{equation}
\frac{dr}{dt}=\pm \left( 1-\frac{r^{2}}{r_{c}^{2}}\right) \sqrt{\frac{%
r^{2}-r_{0}^{2}}{r_{c}^{2}-r_{0}^{2}}},  \label{30}
\end{equation}%
which directly results from the Lagrangian in Eq. (\ref{22}) when $\dot{r}%
^{2}/\dot{t}^{2}=\left( dr/dt\right) ^{2}$. It can be shown that the
solution to this differential equation is 
\begin{equation}
r=\frac{r_{0}}{\sqrt{1-\left( 1-\frac{r_{0}^{2}}{r_{c}^{2}}\right) \tanh
^{2}\left( \pm \frac{t}{r_{c}}\right) }}  \label{31}
\end{equation}%
where we set $t_{0}=0.$ Note that as $t\rightarrow \infty $, the radial
coordinate approaches $r_{c}$, so the distant observer never sees the
particle crossing the cosmic horizon in the de Sitter spacetime. However, in
the presence of a cosmic gate at $r=R_{0}$, the distant observer witnesses
the transition of the particle through the gate at 
\begin{equation}
t_{R}=r_{c}\tanh ^{-1}\left( \sqrt{\frac{1-r_{0}^{2}/R_{0}^{2}}{%
1-r_{0}^{2}/r_{c}^{2}}}\right) .  \label{32}
\end{equation}%
\begin{figure}[tbp]
\includegraphics[width=70mm,scale=1]{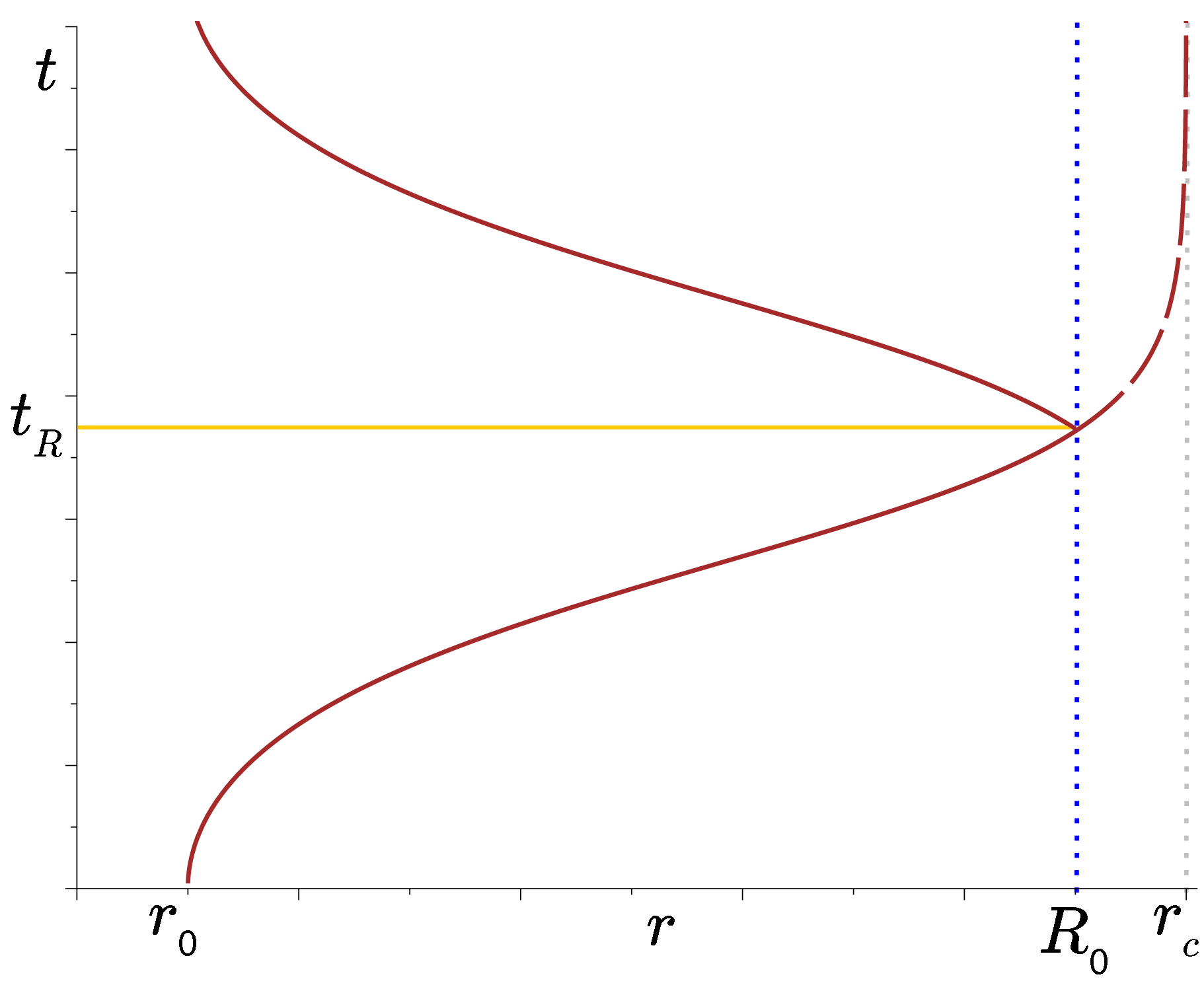}
\caption{Radial timelike-geodesics for a massive particle as seen by a
distant observer.}
\end{figure}

In Fig. 3, we plotted the time $t$ versus radial coordinates $r$ for a
massive particle as observed by a distant observer in the presence (solid
line) and absence (dash line) of the cosmic gate. As it is seen, the
particle's geodesics is continuous across the cosmic gate, although it
undergoes an impulse in colliding with the matter at the gate.

\subsection{The Null Geodesics}

From (\ref{24}) and (\ref{26}), with $L=0$ and $\epsilon =0$ for a photon
(null particle), one finds 
\begin{equation}
\dot{r}^{2}=E^{2}=\left( 1-\frac{r^{2}}{\ell ^{2}}\right) ^{2}\left( \frac{dr%
}{dt}\right) ^{2}  \label{33}
\end{equation}%
for a radial motion, which admits the integral ($t_{0}=0$) 
\begin{equation}
r=r_{c}\left( \frac{r_{0}+r_{c}\tanh \left( \pm \frac{t}{r_{c}}\right) }{%
r_{c}+r_{0}\tanh \left( \pm \frac{t}{r_{c}}\right) }\right)  \label{34}
\end{equation}%
between the radial coordinate and the time $t$ measured by a distant
observer. It could be easily checked that the distant observer never detects
the light reaching the cosmic horizon, although it could pass over the
cosmic gate at a finite time given by 
\begin{equation}
t=r_{c}\tanh ^{-1}\left( \frac{r_{c}\left( R_{0}-r_{0}\right) }{%
r_{c}^{2}-R_{0}r_{0}}\right)  \label{35}
\end{equation}%
if it existed. The geodesics of the massless particle are plotted in Fig. 4. 
\begin{figure}[tbp]
\includegraphics[width=70mm,scale=1]{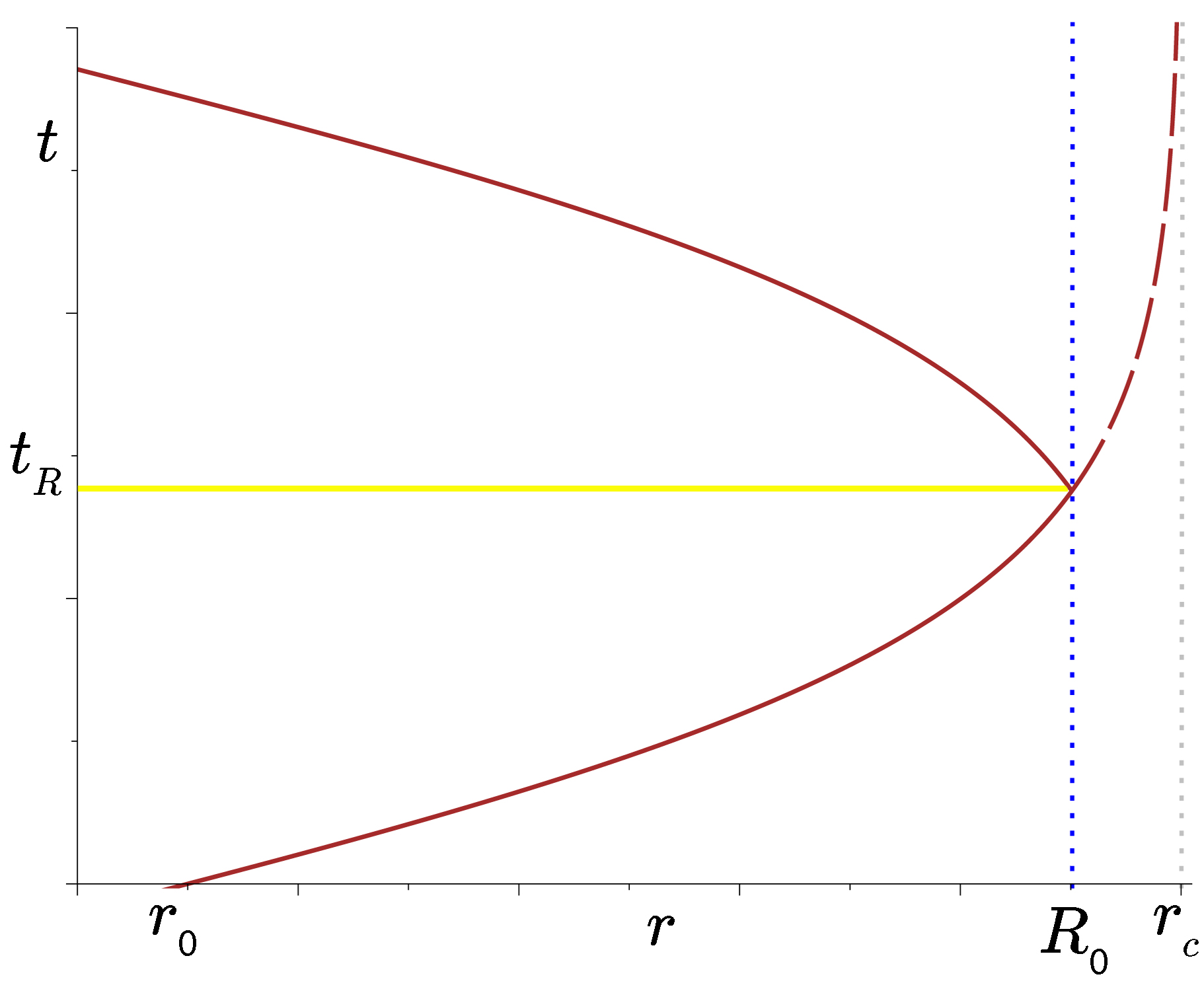}
\caption{Radial null-geodesics for a massless particle as seen by a distant
observer.}
\end{figure}

\section{Conclusion}

In contrast with the concept of a wormhole that connects two different
distant points of the same or different universe(s) through a
hyperplane/throat whose surface area is a local minimum, we introduced a
cosmic gate that connects two closed universes through a hyperplane/gate
whose surface area is a local maximum. In particular, we applied the
junction formalism and glued the regular part of two de-Sitter spacetimes at
a hyperplane located before the cosmological horizon. The resultant universe
in its embedded diagram displayed in Fig. 1 is a closed universe in the
sense that the radial coordinate is confined to a finite interval. We
investigated the mechanical stability of the newly built cosmological object
(we called it \textit{Cosmic Gate}) and showed that it is stable against a
radial linear perturbation. Furthermore, we studied the timelike and null
geodesics of the massive and massless particles in this spacetime.
Specifically, we focused our attention on radial geodesics to show that the
spacetime is geodesically complete. Our results have been presented
analytically and as have been depicted in Fig. 3 and 4, the particle reaches
the gate in a finite time and travels to the other side. As a final remark,
due to the direction of the 4-vector on the gate which is in the opposite
direction of its counterpart on a thin-shell wormhole, the matter presented
at the gate satisfies the physical energy conditions and therefore is normal
matter. This can also be seen from the equal signs of the principal
curvature of the gate (see Fig. 1). We recall that at the throat of a
wormhole, the signs of the two principal curvatures are opposite of each
other indicating the violation of the energy conditions.

\end{document}